\documentclass[prd,twocolumn,letterpaper,superscriptaddress,nofootinbib]{revtex4}
\usepackage{amsmath,epsfig,natbib,bm,psfrag}

\begin{document}
\newcommand{\numfrac}[2]{{\textstyle \frac{#1}{#2}}}
\newcommand\ba{\begin{eqnarray}}
\newcommand\ea{\end{eqnarray}}
\newcommand\be{\begin{equation}}
\newcommand\ee{\end{equation}}
\newcommand\lagrange{{\cal L}}
\newcommand\cll{{\cal L}}
\newcommand\clx{{\cal X}}
\newcommand\clz{{\cal Z}}
\newcommand\clv{{\cal V}}
\newcommand\clo{{\cal O}}
\newcommand\cla{{\cal A}}
\newcommand{\grad}{\nabla}
\newcommand{\Gauss}{\text{G}}

\newcommand{\Psil}{\Psi_l}

\newcommand{\bsigma}{{\bar{\sigma}}}
\newcommand{\bI}{\bar{I}}
\newcommand{\bq}{\bar{q}}
\newcommand{\bv}{\bar{v}}

\newcommand\del{\nabla}
\newcommand\Tr{{\rm Tr}}
\newcommand\half{{\frac{1}{2}}}
\newcommand\bibi{\bibitem}

\newcommand\calS{{\cal S}}
\renewcommand\H{{\cal H}}
\newcommand\K{{\cal K}}
\newcommand\opacity{\tau_c^{-1}}

\renewcommand\P{{\cal P}}

\newcommand{\calE}{{\cal E}}
\newcommand{\calB}{{\cal B}}

\newcommand{\la}{\langle}
\newcommand{\ra}{\rangle}
\newcommand{\kf}{\beta}
\newcommand{\kfprod}{\alpha}

\newcommand{\Omtot}{\Omega_{\mathrm{tot}}}
\newcommand\xx{\mbox{\boldmath $x$}}
\newcommand{\phpr} {\phi'}
\newcommand{\gam}{\gamma_{ij}}
\newcommand{\sqgam}{\sqrt{\gamma}}
\newcommand{\delk}{\Delta+3{\K}}
\newcommand{\dph}{\delta\phi}
\newcommand{\om} {\Omega}
\newcommand{\dom}{\delta^{(3)}\left(\Omega\right)}
\newcommand{\rar}{\rightarrow}
\newcommand{\Rar}{\Rightarrow}
\newcommand{\labeq}[1] {\label{eq:#1}}
\newcommand{\eqn}[1] {(\ref{eq:#1})}
\newcommand{\labfig}[1] {\label{fig:#1}}
\newcommand{\fig}[1] {\ref{fig:#1}}
\newcommand\gsim{ \lower .75ex \hbox{$\sim$} \llap{\raise .27ex \hbox{$>$}} }
\newcommand\lsim{ \lower .75ex \hbox{$\sim$} \llap{\raise .27ex \hbox{$<$}} }
\newcommand\bigdot[1] {\stackrel{\mbox{{\huge .}}}{#1}}
\newcommand\bigddot[1] {\stackrel{\mbox{{\huge ..}}}{#1}}
\newcommand{\Mpc}{\text{Mpc}}

\newcommand{\curl}{\,\mbox{curl}\,}
\newcommand{\ord}{\mbox{O}}
\newcommand{\sigt}{\sigma_{\mathrm{T}}}
\newcommand{\nelec}{n_{\mathrm{e}}}
\newcommand{\ud}{{\mathrm{d}}}
\newcommand{\uD}{{\mathrm{D}}}
\newcommand{\tc}{t_{\mathrm{c}}}
\newcommand{\clq}{{\mathcal{Q}}}
\newcommand{\TT}{{\mathrm{TT}}}
\newcommand{\clh}{{\mathcal{H}}}
\newcommand{\clp}{{\mathcal{P}}}

\newcommand{\nonflat}[1]{}

\title{CMB anisotropies from primordial inhomogeneous magnetic fields}

\author{Antony Lewis}
 \email{antony@cosmologist.info}
 \affiliation{CITA, 60 St. George St, Toronto M5S 3H8, ON, Canada.}

\begin{abstract}
Primordial inhomogeneous magnetic fields of the right strength can leave a signature on the
CMB temperature anisotropy and polarization. Potentially observable
contributions to polarization B-modes are generated by 
vorticity and gravitational waves sourced by the magnetic anisotropic stress. We compute the corresponding CMB transfer functions in
detail including the effect of neutrinos. The shear rapidly causes the
neutrino anisotropic
stress to cancel the stress from the magnetic
field, suppressing the production of gravitational waves and vorticity
on super-horizon scales after neutrino decoupling. A significant large
scale signal from tensor modes can only be produced before neutrino
decoupling, and the actual amplitude is somewhat uncertain. Plausible
values suggest primordial nearly scale invariant fields with $B_\lambda\sim 10^{-10}\Gauss$ today may be observable from their large scale
tensor anisotropy. They can be distinguished from primordial
gravitational waves by their non-Gaussianity.
Vector mode vorticity sources B-mode power on much smaller scales
with a power spectrum somewhat similar to that expected from weak
lensing, suggesting amplitudes $B_\lambda\sim 10^{-9}\Gauss$ may be
observable on small scales for a spectral index $n\sim -2.9$. In the appendix we review the covariant equations
for computing the vector and tensor CMB power spectra that we implement numerically.
\end{abstract}
\maketitle

\section{Introduction}

Magnetic fields are ubiquitous in the universe, with  $\sim
10^{-6}\Gauss$ coherent fields observed on galactic and
cluster scales. However their origin
is not well understood (for a review see Ref.~\cite{Grasso:2000wj}). Tiny seed fields $\alt 10^{-20}\Gauss$ may
have been amplified by a dynamo
mechanism to give the much larger fields we now see,
though to what extent this process can work in practice is not yet clear~\cite{Brandenburg:2004jv}.
Alternatively initial fields of strength $\sim10^{-9}\Gauss$
can give rise to galactic fields of the observed values without
a functioning dynamo mechanism. Such fields have potentially interesting
observational signatures on the CMB, and if present would provide powerful
constraints on models of the early universe. The absence of such signatures
may also serve as a consistency check on models of galaxy evolution that would be
observationally incompatible with initial fields this large.

A primordial field of $\sim10^{-9}\Gauss$ can leave a signature in the B-mode
(curl-like) CMB polarization. Since the scalar (density)
perturbations do not produce B-modes at linear order, the B-modes are
a much cleaner signal of additional physics than very small fractional
changes to the temperature or E-mode polarization. However B-modes are
produced at second order through weak lensing~\cite{Zaldarriaga98,Hu00}, and are also generated
by primordial gravitational waves (tensor modes)~\cite{Seljak:1997gy}.
 Other possible sources include
topological defects~\cite{Seljak:1997ii,Pogosian:2003mz}. The focus of many future CMB observations
will be on observing the B-modes, so it is useful to assess in detail
the various possible components and how they can be distinguished from
each other. 

Primordial fields with a blue spectrum
compatible with nucleosynthesis are far too weak on cosmological
scales to leave an interesting signature~\cite{Caprini:2001nb}.
In this paper we consider in detail the CMB signal expected from $\sim
10^{-9}\Gauss$ primordial fields with a nearly scale
invariant spectrum.
Such fields are not well motivated by current
theoretical models, which mostly give much smaller amplitudes or a much
bluer
spectrum~\cite{Grasso:2000wj,Durrer:2003ja,Enqvist:2004yy,Turner:1988bw}.
Observation of primordial fields at this level would therefore be a powerful
way to rule out many models. However Ref.~\cite{Bamba:2003av} present
one model in which observably interesting CMB signatures may be produced.

 Previous semi-analytical work has
investigated the CMB signal from both tensor~\cite{Durrer:1999bk,Mack:2001gc} and vector modes~\cite{Mack:2001gc,Subramanian:2003sh} sourced by
magnetic fields. Here we give a more detailed numerical analysis of
the full linearized equations. We include the effect of neutrinos as
they change the way
that magnetic fields source gravitational waves and vorticity on super-horizon
scales. Our final CMB power spectra include the contribution to the
B-mode signal from both the tensor and vector
modes.
We do not consider helical
modes~\cite{Pogosian:2002dq,Caprini:2003vc} which can be detected via
their parity-violating correlations, nor the effects of Faraday
rotation~\cite{Kosowsky:1996yc,Scoccola:2004ke} (which can be identified
by the frequency dependence). We also assume that reionization is
relatively sharp and unmodified by energy injection into the IGM from
decay of the small scale magnetic
field~\cite{Sethi:2004pey}. For a discussion of constraints on
homogeneous magnetic fields see e.g. Ref.~\cite{Chen:2004nf} and
references therein.

\section{Covariant equations}

We consider linear perturbations in a flat FRW universe evolving
according to general relativity with a cosmological constant and cold
dark matter, and approximate the neutrinos as massless. Perturbations
can be described covariantly in terms of a 3+1 decomposition with
respect to some choice of observer velocity $u_a$ (we use natural
units, and the signature where $u_a u^a=1$), following Refs.~\cite{Ellis83,Gebbie99,Challinor:1998xk,Tsagas:1998jm}.
The stress-energy tensor can be decomposed with respect to $u_a$ as
\ba
T_{ab} = \rho u_a u_b - p h_{ab} + 2 u_{(a} q_{b)} + \pi_{ab}
\ea
where $\rho$ is the energy density, $p$ is the pressure, $q_a$ is the
heat flux and  and $\pi_{ab} \equiv
T_{\la ab\ra}$ is the anisotropic stress. Angle brackets around
indices denote the projected (orthogonal to
$u_a$) symmetric trace-tree part (PSTF).
The tensor
\ba
h_{ij} \equiv g_{ij} -u_i u_j,
\ea
where $g_{ij}$ is the metric tensor,
projects into the instantaneous rest space orthogonal to $u_a$. It defines a
spatial derivative $\uD_a \equiv h_a{}^b \grad_b$ orthogonal to $u_a$ where $\grad_a$ is
the covariant derivative.
Spatial derivatives can be used to
quantify perturbations to background quantities, for example the pressure
perturbation can be described covariantly in terms of $\uD_a p$. 

Conservation of total stress-energy $\grad^a T_{ab}=0$ implies an
evolution equation for the total heat flux $q_a$
\ba
\dot{q}_a + \frac{4}{3} \Theta q_a + (\rho+p)A_a - \uD_a p + \uD^b \pi_{ab}= 0
\ea
where $\dot{q}_a \equiv u^b\grad_b q_a$,
$\Theta \equiv \grad^a u_a$ is three times the Hubble expansion, and 
$A_a \equiv u_b\grad^b u_a$ is the acceleration. 
The evolution equation for the heat flux $q_a^i = (\rho^i+p^i) v_a^i$ of each
matter component present is of the form
\ba
\dot{q}_a^i + \frac{4}{3} \Theta q_a^i + (\rho^i+p^i)A_a - \uD_a p^i + \uD^b \pi_{ab}^i= L_a^i
\label{qevolve}
\ea
where $L_a^i$ is an interaction force term. Conservation of total
stress energy $q_a = \sum_i q^i_a$ implies that $\sum_i L_a^i=0$.
For magnetic fields the components of the stress-energy tensor are given
by\footnote{Note that unlike many other authors we use natural rather
  than Gaussian units. Due to the signature choice $-E^2\ge 0$.}
\ba
\rho^B &=& 3p^B = -\half (E^2 + B^2) \\
\pi_{ab}^B &=& - E_{\la a} E_{b\ra}- B_{\la a} B_{b\ra} \\
 q_a^B &=&
-(E \times B)_a
\ea
where $E_a$ and $B_a$ are the electric and magnetic field projected vectors. We take $B^2$ and $E^2$ to be
first order, and to this order $E_a$ and $B_a$ are frame invariant. For most of its
evolution the universe is a good conductor so we may take $E_a=0$ in
all linear frames: the magnetic fields are frozen in, and in this approximation almost all the complications
of MHD disappear. The linearized Bianchi identity for the electromagnetic field tensor implies
\ba
\dot{B}_a + \frac{2}{3}\Theta B_a = 0
\ea
so the magnetic field simply redshifts as $1/S^2$ where $S$ is the
scale factor, and hence $\pi^B_{ab}
\propto \rho^B \propto 1/S^4$.
More general equations can be found in e.g.~Ref.~\cite{Tsagas:1998jm}.

The Poynting vector heat flux is zero
($q^B_a=0$) in all linear frames because we have set $E_a=0$. Since $A_a$ is first order,
on linearizing we have the Lorentz force $L^B_a$ given by the
evolution equation~\eqref{qevolve}
\be
 \uD^b \pi^B_{ab}- \uD_a p^B = L_a^B.
\ee
This is consistent with the usual $\curl B \times B$ expression. 
The opposite force acts on the baryons to ensure total momentum
conservation, which with the Thomson
scattering terms~\cite{Challinor:1998xk} gives the baryon velocity evolution equation:
\begin{multline}
\dot{v}_a + \frac{1}{3}\Theta v_a + A_a - \frac{D_a p^b}{\rho^b}= \\-\frac{\rho^\gamma}{\rho^b} \left[ n_e \sigma_T \left(
  \frac{4}{3}v_a - I_a\right) + \frac{\uD^b \pi_{ab}^B- \uD_a p^B}{\rho^\gamma}  \right]
\label{vev}
\end{multline}
where $n_e$ is the electron number
density, $\sigma_T$ is the Thomson scattering cross-section,
$\rho^\gamma$ is the photon energy density, and we
neglect baryon pressure terms of the form
${p^b}'/{\rho^b}'\ll 1$.
Thus magnetic fields source baryon vorticity, as well as providing
extra density and pressure perturbations, and anisotropic stresses.

We define the vorticity vector $\Omega_a \equiv  \curl u_a$
where for a general tensor
\ba
\curl X_{a_1\dots a_l} \equiv \eta_{bcd(a_1} u^b \uD^c X^{d}{}_{a_2\dots a_l)}
\ea 
and round brackets denote symmetrization. It is transverse $\uD^a \Omega_a = 0$.
Remaining quantities we shall need are\footnote{From here on we do not
  use $E$ for the electric field; $E_{ab}$ has nothing to do with
  electromagnetism, and is merely called the analogous `electric' part of the
  Weyl tensor by analogy. The `electric' part of the polarization
  distribution is
  written as $\calE_{A_l}$.} the `electric' $E_{ab}$ and
`magnetic' $H_{ab}$ parts of the Weyl tensor $C_{abcd}$
\ba
E_{ab} \equiv C_{acbd} u^c u^d \quad\quad 
H_{ab} \equiv  \frac{1}{2} \eta_{acdf} C_{be}{}^{cd}u^e u^f
\ea
(which are frame invariant) and the shear $\sigma_{ab} \equiv \uD_{\la a} u_{b \ra}$. The Einstein
equation and the Bianchi identity give the constraint equations
\begin{eqnarray}
\uD^a \sigma_{ab} - \frac{1}{2}\curl \Omega_b - {\frac{2}{3}}\uD_b \Theta - \kappa q_b
= 0 \nonumber\\
\uD^a E_{ab} - \kappa \left({\frac{\Theta}{3}} q_b + {\frac{1}{3}}
\uD_b \rho + {\frac{1}{2}} \uD^a \pi_{ab}\right) =0 \nonumber\\
\uD^a H_{ab} - {\frac{1}{2}}\kappa [ (\rho + p)\Omega_b + \curl q_b ]=0
\nonumber \\
H_{ab} - \curl \sigma_{ab} + \frac{1}{2}\uD_{\langle a} \Omega_{b \rangle}=0,
\label{constraints}
\end{eqnarray}
and the evolution equations
\begin{eqnarray}
\dot{\Omega}_a + \frac{2}{3}\Theta\Omega_a &=& \curl A_a  \nonumber\\
\dot{\sigma}_{ab} + \frac{2}{3} \Theta \sigma_{ab} & = &  - E_{ab}
- {\frac{1}{2}}\kappa \pi_{ab} \label{eq:tp3} \nonumber\\
\dot{E}_{ab}+ \Theta E_{ab} & = & \curl H_{ab} + {\frac{\kappa}{2}}
\left[  \dot{\pi}_{ab}  -(\rho + p)\sigma_{ab}
+ {\frac{\Theta}{3}}\pi_{ab}\right] \label{eq:tp4} \nonumber\\
\dot{H}_{ab} +\Theta H_{ab}& = & - \curl E_{ab} -
{\frac{\kappa}{2}}\curl \pi_{ab}.
\label{props}
\end{eqnarray}
Here $\kappa \equiv 8\pi G$.

The distribution functions for the various species can be expanded
into multipole moments. For example the photon
multipole tensors $I_{A_l}\equiv I_{\la a_1\dots a_l \ra}$ are defined as
moments of the distribution of the photon intensity $I(e)$ per solid
angle as~\cite{Challinor:1999xz}
\be
I_{A_l} \equiv  \int d\Omega_e\, I(e)\,\,e_{\la A_l \ra},
\ee
where the direction vector $e_a$ is normalized to $e^a e_a =-1$ and
$e_{\la A_l\ra} = e_{\la a_1}\dots e_{a_l\ra}$ are irreducible PSTF
tensors. The $e_{\la A_l\ra}$ are orthogonal:
\be
\frac{1}{4\pi} \int d\Omega_e\, e_{\la A_l\ra} e^{\la B_n \ra} =
\delta_{ln} \frac{(-2)^l (l!)^2}{(2l+1)!} h^{\la b_1}_{\la a_1}\dots h^{b_l\ra}_{a_l\ra}.
\ee
The $I_{A_l}$ multipole tensors have $2l+1$ degrees of freedom, $I_{ab}=\pi_{ab}^\gamma$ is the anisotropic
stress, $I_{a}=q_a^\gamma$ is the heat flux and $I=\rho_\gamma$. 
The temperature anisotropy can then be expanded as
\be
\frac{\Delta T(e)}{T} = 
\sum_l \frac{ (2l+1)! }{ 4(-2)^l (l!)^2 } \frac{ I_{A_l} e^{A_l} }
{\rho_\gamma} = \sum_{l} \sum_{m=-l}^l a_{lm} Y_{lm}(e)
\ee
where the latter expansion in terms of spherical harmonics $Y_{lm}$ is
the non-covariant version of the expansion in $I_{A_l}$.
The CMB power spectrum is defined in terms of the variance of the
spherical harmonic components $a_{lm}$ by 
\be
C_l \equiv \la |a_{lm}|^2\ra = \frac{\pi}{4} \frac{(2l)!}{(-2)^l(l!)^2} \frac{\la I_{A_l} I^{A_l} \ra}{\rho_\gamma^2}.
\label{C_l}
\ee
Analogous results for the polarization are given in
Ref.~\cite{Challinor:2000as}, where $\calE_{A_l}$ is a gradient-like
multipole of the polarization tensor and $\calB_{A_l}$ is a curl-like
multipole. 

The covariant equations can be expanded in terms of scalar, vector and
tensor harmonics. The details and definitions are given in
the appendix. In the following sections we analyse in detail the
tensor and vector equations, where each quantity is a component of a
harmonic expansion, and $k$ labels are suppressed. Scalar modes can
source temperature and E-polarization CMB signals, however since we
are mostly interested in the B-polarization signal, which is not
sourced by scalar modes, we do not discuss scalar modes here. A
partial analysis of scalar modes is given in Ref.~\cite{Koh:2000qw}.

\section{Tensors}

Expanding in $m=2$ tensor harmonics~\eqref{Qexpand} (and suppressing
$m$ indices), the constraint equations~\eqref{constraints} imply that
the Weyl tensor variable $H$ is related to the shear by $H = \sigma$.
The evolution equations~\eqref{props} then give
\ba
k^2(E' + \H E) - k^3 \sigma + \frac{\kappa}{2} S^2(\rho+p) k \sigma &=&
\frac{\kappa}{2} S^2(\Pi' + \H \Pi)\nonumber\\
\sigma' + 2\H \sigma + k E &=& - \frac{\kappa S^2 \Pi}{2k}
\ea
where primes denote derivatives with respect to conformal time $\eta$ and $\H\equiv
S\Theta/3$ is the conformal Hubble parameter. 
The Weyl tensor variable
$E$ and the shear $\sigma$ define the new variable
\be
H_T \equiv -2 E - \frac{\sigma'}{k}
\ee
to correspond to the metric
perturbation variable of non-covariant approaches. It satisfies 
$H_T' = -k\sigma$, and the above equations
combine to give the well known evolution equation
\be
H_T'' + 2\H H_T' + k^2 H_T = \kappa S^2 \Pi.
\ee
Magnetic fields provide a component of the anisotropic stress $\Pi$
and hence source gravitational waves, and we quantify the magnetic field source by the
  dimensionless ratio $B_0 \equiv \Pi_B/\rho_\gamma$.
The covariant tensor equations
are discussed in more detail in
Ref.~\cite{Challinor:1999xz}. 

 Equations for the evolution of the tensor
multipoles are obtained from the appendix ($m=2$ in Eqs.~\eqref{Imult}
  and ~\eqref{EBmult}). We use a series expansion in conformal time
  $\eta$ to identify the
regular primordial modes in the early radiation dominated era.
Defining $\omega\equiv
\Omega_m\H_0/\sqrt{\Omega_R}$, where $\Omega_R =
\Omega_\gamma+\Omega_\nu$, and $\H_0$ and $\Omega_i$ are the Hubble
parameter and densities (in units of the critical density) today,
the Friedmann equation gives
\ba
S = \frac{\Omega_m \H_0^2}{\omega^2}\left( \omega\eta +
\frac{1}{4}\omega^2\eta^2 +\clo(\eta^5) \right).
\ea
Defining the ratios $R_\nu\equiv\Omega_\nu/\Omega_R$, $R_\gamma
\equiv\Omega_\gamma/\Omega_R$
and keeping lowest order terms the regular solution (with zero initial
anisotropies for $l>2$) is
\ba
H_T &=&  H_T^{(0)}\left( 1  -\frac{5}{2} \frac{(k\eta)^2}{4R_\nu+15}\right)+ \frac{15}{28}  \frac{R_\gamma B_0 (k\eta)^2 }{4R_\nu+15}\nonumber\\
\sigma &=& \frac{5 H_T^{(0)} k\eta}{4R_\nu+15}  -\frac{15}{14}  \frac{R_\gamma B_0 k\eta }{4R_\nu+15} \\
\pi_\nu &=& - \frac{R_\gamma B_0}{R_\nu} \left( 1 - \frac{15}{14} \frac{(k\eta)^2}{4R_\nu+15}\right) + \frac{4}{3} \frac{(k\eta)^2
  H_T^{(0)}}{4R_\nu+15}\nonumber
\ea
where $H_T^{(0)}$ is the initial value (after neutrino decoupling) and
$\pi_\nu \equiv \Pi_\nu/\rho_\nu$.
The $B_0 \ne 0$ mode (with  $H_T^{(0)}=0$) has compensating anisotropic stresses: the
sum of the magnetic and neutrino terms gives the total source term
\be
 \kappa S^2 \Pi 
=  \frac{45}{14} \frac{R_\gamma k^2 B_0}{4R_\nu+15}
 \left( 1 - \frac{(45 - 2 R_\nu)\omega\eta}{2R_\nu + 15}\right) + {\cal O}(\eta^2)
\ee
rather than the $\propto S^2\rho_\gamma \propto 1/\eta^2$ result expected without
collisionless radiation. For $k \ll \H$ there is therefore no
sourcing of gravitational waves during radiation domination, so $H_T
 \propto (k\eta)^2$ if it was zero initially. Collisionless fluids suppress generation of
gravitational waves on super-horizon scales.

Before neutrino decoupling there is no neutrino
anisotropic stress so the magnetic field source is not compensated.
Taking $\eta_{in}$ as the magnetic field production time (at which we
take $H_T=\sigma=0$), 
the solution for $k\eta \ll 1$ is approximately\footnote{
It appears that $\sigma$ will
  be large when $k\eta \ll 1$, however the
  contribution to $\la \sigma_{ab} \sigma^{ab}\ra/(\kappa\rho_\gamma)$
  and similar terms remain
  small,  so this does not violate the linearity assumption.
}
\ba
H_T &\approx& 3R_\gamma B_0\left(\ln(\eta/\eta_\text{in}) +
  \frac{\eta_{in}}{\eta} -1\right)\\
\sigma &\approx&-\frac{3R_\gamma B_0}{k\eta}\left(1 - \frac{\eta_{in}}{\eta}\right). 
\ea
After neutrino decoupling this mode must convert into a combination of
the above regular modes, which can then be used to compute the
observable signature. As the neutrino coupling is switched off the 
neutrino anisotropic
stress becomes important, and with no scattering evolves as
\be
\pi_\nu' = -\frac{k}{3} J_3 + \frac{8}{15} k\sigma.
\ee
Since the octopole $J_3$ and $\pi_\nu$ will be zero before neutrino decoupling, for
modes well outside the horizon we have
$\pi_\nu' \sim -B_0/\eta$ just after decoupling (we assume $\eta\gg \eta_{in}$), and hence
the neutrino anisotropic stress grows logarithmically $\pi_\nu \sim -B_0\ln
(\eta/\eta_*^\nu)$. It therefore reaches the constant value $\pi_\nu \sim -B_0$
of the regular solution in about one e-folding. At this point $H_T$
ceases to grow logarithmically because the magnetic anisotropic stress
source is now cancelled by the neutrinos, and $H_T$ gives the amplitude of the
usual regular solution $H_T^{(0)}$. 

Thus after neutrino decoupling we expect a combination of the usual
passive primordial tensor mode and the regular compensated sourced
mode. As we show explicitly below, the compensated mode can be
neglected compared to the small scale vector mode
contribution. The passive tensor mode has
\be
H_T^{(0)} \approx 3R_\gamma B_0\ln(\eta_*^\nu/\eta_\text{in})
\ee
where $\eta_*^\nu$ is the time of neutrino decoupling (assuming magnetic
field generation is during radiation domination at
$\eta\sim\eta_\text{in}$ and before neutrino
decoupling). Since the transverse traceless part of the metric tensor $h_{ij} = \sum_{k,\pm} 2 H_T Q^2_{ij}$ this corresponds to a
primordial power spectrum for $h$ of
\be
P_h \approx \left[6R_\gamma \ln(\eta_*^\nu/\eta_\text{in})\right]^2 P_{B_0}.
\ee
Power spectra are defined in Eq.~\eqref{Pdef} and $R_\gamma \sim 0.6$.

Thus the tensor covariance part of the magnetic field signal is very
similar to that expected from primordial gravitational waves, and can
be computed trivially by using the above tensor power spectrum in the
numerical codes CAMB~\cite{Lewis:1999bs} or CMBFAST~\cite{Seljak:1996is}. 
However unlike primordial gravitational waves the spectral index of
$P_h$ here is expected to be at least
slightly blue, and the signal should be non-Gaussian because
$B_0$ is quadratic in the magnetic field. Allowing for subtracting the
B-mode lensing signal, levels of $P_h \sim 10^{-15}$ may be
observable\footnote{Observable in the sense that in the null hypothesis that there are no
non-lensing B-modes, the residual noise level would be consistent with
this level. Actual subtraction of CMB lensing in the presence of
B-modes from magnetic field sourced vector modes may be extremely
difficult, but one should still be able to detect violations of the null
hypothesis that there are none.}~\cite{Seljak:2003pn}.
 This corresponds to $B_\lambda \sim 10^{-10} \Gauss$ for
$\eta_\text{in}/\eta_*^\nu \sim 10^{-6}$ and a close
to scale invariant spectrum.
To distinguish this from primordial gravitational waves from
inflation one would need to detect the non-Gaussianity or small scale
power from the vector modes.

The compensation mechanism in principle works with any collisionless
relativistic fluid, even when it only makes up a fraction
$R_\nu\rightarrow 0$ of the energy density ($R_\nu$ can be interpreted
as any collisionless component).
However
if there is collisionless relativistic matter only from a time
$\eta_*$ after magnetic field production, partial compensation will be effective at a time $\eta
\sim \eta_* e^{R_\gamma/R_\nu}$. For small fractions $R_\nu$ this is a very large
time, so the mechanism is inefficient for components
that are only a small fraction of the density allowed by the
nucleosynthesis bound. Neutrinos themselves can only suppress
gravitational wave production after neutrino decoupling at $z\sim 10^9$.

\section{Vectors}


Unlike tensors modes, vectors are not in general frame invariant.
It is therefore convenient to choose the frame $u_a$ to simplify the analysis. 
At linear order one can
always write $u_a = u_a^\perp +
v_a$, where $u_a^\perp$ is hypersurface orthogonal and $v_a$ is first order, so $\curl u_a =
\curl v_a$. For a zero order scalar quantity $X$ it follows that 
\be
\uD_a X =
\uD_a^{\perp} X - v_a \dot{X}.
\ee 
For vector modes $(\uD_a^{\perp} X)^{(1)} =0$, and 
it is convenient to choose the frame $u_a$ to be
hypersurface orthogonal so that $\curl u_a = 0$ and hence $(\bar{\uD}_a
X)^{(1)}=0$, 
where the bar denotes evaluation in the zero vorticity frame.
From its propagation equation, vanishing of the vorticity also implies that $\bar{A}_a^{(1)}=0$,
so the zero vorticity frame coincides with the synchronous
gauge. The CDM velocity is also zero in this frame modulo a mode that decays as
$1/S$ where $S$ is the scale factor. 

Expanding in the $m=1$ vector harmonics~\eqref{Qexpand}
the equations for the harmonic coefficients in the zero vorticity
frame ($\Omega=0$) reduce to
\ba
k(\bsigma' + 2\H \bsigma) = - \kappa S^2\Pi \nonumber\\
H = \half\bsigma \quad\quad\quad 2\kappa S^2 \bq = k^2\bsigma.
\ea
The combination $v+\sigma$ (the Newtonian gauge velocity) is frame invariant, as are
$\bsigma = \sigma + \Omega$ and $\bv = v - \Omega$. By choosing to
consider the
zero vorticity frame we have simply expedited the derivation of the
above frame invariant equations. Other papers use
the Newtonian gauge~\cite{Hu:1997hp}, in which $\bsigma$ is
the vorticity. The equations are consistent.
The baryon vorticity evolution equation~\eqref{vev} becomes
\ba
\bv' + \H \bv = -\frac{\rho_\gamma}{\rho_b} \left[ S n_e \sigma_T \left(
  \frac{4}{3}v - I_1\right) + \half k B_0 \right]
\ea
where $I_1 = 4v_\gamma/3$ and as before $B_0 \equiv
\Pi_B/\rho_\gamma$.

At early times the baryons and photons are tightly coupled, the
opacity $\opacity \equiv S \text{n}_e \sigma_{\text{T}}$ is
large. This means $v_\gamma \approx  v$, and we
can do an expansion in $\tau_c$ that is valid for
$\epsilon\equiv \max(k\tau_c, \H\tau_c) \ll 1$. To lowest order the
baryon velocity evolves as
\ba
\bv' &=& -\frac{R}{1+R}\left ( \H \bv + \frac{3}{8} \frac{k B_0}{R}\right)
+{\cal O}(\tau_c)
\ea
where $R\equiv 3\rho_b/4\rho_\gamma$.
The solution with zero initial vorticity is
\ba
\bv \approx -\frac{3}{8} \frac{B_0 k\eta}{1+R},
\ea
so the magnetic field sources a growing baryon vorticity.  At matter
domination the vorticity starts to redshift away, however by
recombination this will only be an order unity effect.
On smaller scales where $k\tau_c ={\cal O}(1)$ before decoupling the
perturbations are damped by photon diffusion, giving a characteristic
fall off in perturbation power on small scales.


To identify the primordial regular mode we perform a series expansion
as we did for the tensor case.
Assuming no primordial radiation vorticity (the regular vector mode~\cite{Lewis:2004kg}) the result is
\ba
\bsigma &=&  - \frac{45}{14} k\eta \frac{R_\gamma B_0}{4R_\nu+15} \\
\bq_\gamma &=&   - \frac{k\eta B_0}{2}
\\
\bq_\nu &=& \frac{k\eta B_0}{2} \frac{R_\gamma}{R_\nu} \\
\pi_\nu &=& -\frac{R_\gamma}{R_\nu} B_0. 
\ea
As in the tensor case the anisotropic stresses compensate on
super-horizon scales, so they source negligible shear
$\bsigma$ on these scales. However, unlike the tensor case, any non-zero
$\bsigma$ present on super-horizon scales at neutrino decoupling
decays away rapidly and has no observable effect, so the evolution
prior to neutrino decoupling is irrelevant. The observable signature
comes from the vorticity
sourced on sub-horizon scales by the magnetic anisiotropic stress on
its own.

In the approximation that recombination is sharp at $\eta=\eta^*$, the photon
multipoles are given approximately from the integral
solution~\eqref{vecintsol} by
\ba
\frac{I_l(\eta_0)}{4} \approx \left[(\bv+\bsigma)\Psil  + \frac{\zeta}{4} \frac{d\Psil}{d\chi}
\right]_{\eta*} \!\!+   \int_{\eta*}^{\eta_0} d\eta \bsigma' \Psil,
\ea
where $\Psil \equiv l j_l(\chi)/\chi$, $j_l(x)$ is a spherical Bessel
function and $\chi\equiv k(\eta_0-\eta)$. Thus the Newtonian gauge
vorticity 
$\bsigma$ has last scattering and integrated Saches-Wolfe (ISW) contributions to the temperature
anisotropy. In the absence of neutrino anisotropic stress $\bsigma \sim
-3 R_\gamma B_0/(k\eta)$ during radiation domination, as in the tensor case. On super-horizon scales
this  is large, and would give a large scale contribution from $\bsigma$ orders of
magnitude larger than the small super-horizon Doppler contribution
from $\bv$. 
Previous work~\cite{Durrer:1998ya,Mack:2001gc} has neglected the
contributions from $\bsigma$ around last scattering, an approximation
that is good on small scales. However on
super-horizon scales
when $\bsigma$ is large, the diverging $\bsigma$ contributions are far
from negligible. Previous power spectra are approximately the right shape on large scales, but for
the wrong reason: the large scale anisotropies are small because of
neutrino compensation suppressing the source term for $\bsigma$, not
because they are insensitive to $\bsigma$. Similar comments apply to the polarization power
spectra.

There is one caveat to the above. The decay of $\bsigma$ from neutrino
decoupling to last scattering amounts to a decay factor of about
$(z_*^\nu/z_*^\gamma)^2 \sim 10^{12}$. However on arbitrarily large
scales the $1/(k\eta_*^\nu)$ evolution of $\bsigma$ before neutrino decoupling can be larger than this, so on the very largest scales
there can still be a contribution to $\bsigma$ at last scattering. In the
asymptotic limit the dipole $I_1$ appears singular, though the
quadrupole $I_2$ is 
finite. The constraint $n>-3$ ensures that the total power from large
scale modes is not singular if the individual modes are not. Here we
neglect this effect, effectively assuming the power spectrum cuts off
on sufficiently large scales that are otherwise unobservable. It is
unclear whether there is a serious infrared problem with a nearly
scale invariant spectrum or not. We simply compute the CMB transfer
function from a given anisotropy stress $B_0$, and defer the issue of
what the spectrum actually is, how it could be generated, and its actual
asymptotic behaviour.


\section{Numerical Results}

\begin{figure}[t]
\begin{center}
\psfrag{l}[][][1.7]{$\ell$}
\psfrag{Cl}[][][1.7]{$\ell(\ell+1)C_\ell/(2\pi\mu K^2)$}
\psfig{figure=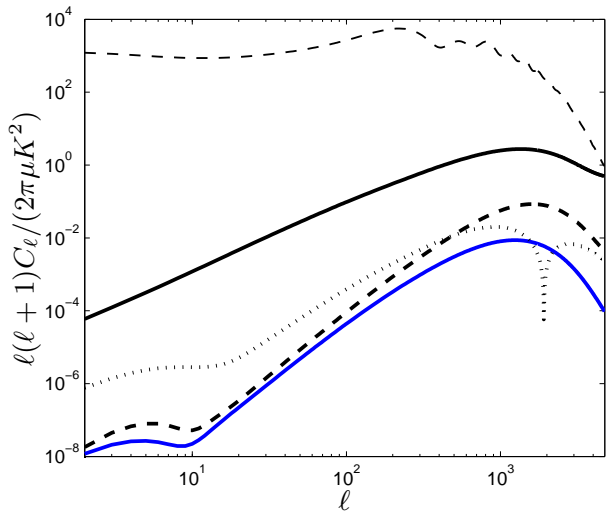,angle=0,width = 8cm}
\caption{
Typical CMB temperature TT (top solid), polarization EE (bottom solid),
BB (dashed thick) and cross-correlation TE (dotted; absolute value) power
spectra from vector modes with $B_\lambda=3\times 10^{-9} \Gauss$, $n=-2.9$. The thin dashed line
shows the scalar adiabatic mode TT spectrum (without magnetic fields). The increase in
power at $\ell\alt 10$ is due to reionization at redshift $z\sim 13$.
\label{vecCls}}
\end{center}
\end{figure}

\begin{figure}[t]
\begin{center}
\psfrag{l}[][][1.7]{$\ell$}
\psfrag{Cl}[][][1.7]{$\ell(\ell+1)C_\ell/(2\pi\mu K^2)$}
\psfig{figure=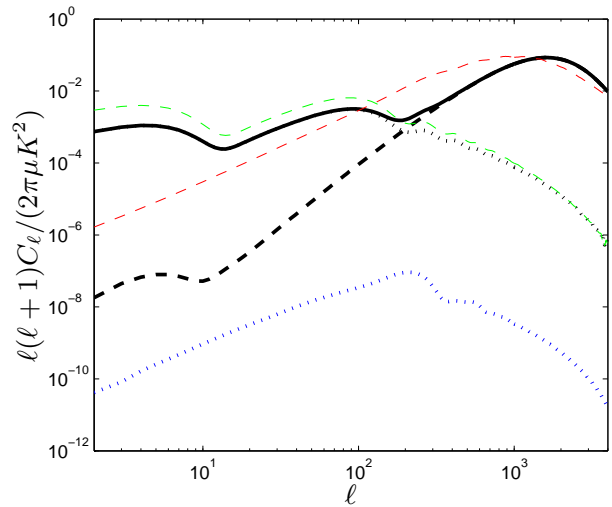,angle=0,width = 8cm}
\caption{
Typical CMB B-mode polarization power spectra
spectra for vector modes (thick dashed), tensor modes (dotted) and
total (solid) for $B_\lambda=3\times 10^{-9} \Gauss$, $n=-2.9$,
$\eta_*^\nu/\eta_{\text{in}}=10^6$. The bottom dotted line is 
from the negligible tensor modes sourced after neutrino decoupling
(the compensated mode). The top dotted
line is from tensors sourced after magnetic field generation until neutrino decoupling (and should be regarded as an estimate correct to
 a few orders of magnitude).
The thin dashed lines
show the  $B$-mode spectrum from weak
lensing (peaking at $\ell\sim 1000$), and scale invariant primordial tensors with initial power
ratio $\sim 10^{-1}$ (peaking at $\ell \sim 100$). The magnetic field
spectra scale as $B_\lambda^4$. 
\label{BCls}}
\end{center}
\end{figure}

In this paper the focus is on calculation of accurate CMB transfer functions
from a given initial distribution. As a convenient ansatz for
computing sample $C_l$  power spectra
we assume a Gaussian primordial magnetic field
distribution, with power spectrum $P_B \propto k^{n-3}$ 
(the definition of $n$ is conventional).

One can define a smoothed
magnetic field $B_\lambda$ using a Gaussian smoothing of width
$\lambda$ (we choose $\lambda = 1 \text{Mpc}$) and express the power
spectrum in terms of $B_\lambda$ as in Refs.~\cite{Mack:2001gc,Durrer:1999bk}.
In harmonic space the anisotropic stress is given by a convolution of
the underlying magnetic fields, so the power spectrum for the
anisotropic stress at a given $k$ feels the power from across the $P_B$ spectrum.
For vectors and tensors the resultant power spectrum is given 
approximately by~\cite{Mack:2001gc}
\begin{multline}
P_{B_0} \approx \frac{4}{(2n+3)}\left[ \frac{(2\pi)^{n+3}
    B_\lambda^2}{2\Gamma(\frac{n+3}{2})\rho_\gamma} \right]^2 \times\\\biggl\{ \left(\frac{k_D}{k_\lambda}\right)^{2n+3}\left( \frac{k}{k_\lambda}\right)^{3} + \frac{n}{n+3} \left( \frac{k}{k_\lambda}\right)^{2n+6}\biggr\}
\end{multline}
for $-3 < n $. The
scale $k_D$ comes from a small scale damping cut-off~\cite{Mack:2001gc,Durrer:1999bk}, and does not
affect the power spectrum significantly for nearly scale invariant
power spectra with $n\sim -3$. The spectrum is singular at $n=-3$,
which comes from the singular build up of super-horizon power for a scale
invariant $B$ spectrum with no large scale cut-off. Since $B_0\equiv \Pi/\rho_\gamma$ is quadratic in
$B$, the spectrum of $B_0$ will be non-Gaussian, so the power spectrum
only contains a subset of the available information, though it is useful
to assess the detectability amplitude.

As a sample example we take $B_\lambda = 3\times 10^{-9}\Gauss$ and
$n=-2.9$ (as in Ref.~\cite{Subramanian:2003sh}), which implies
\be
P_{B_0} \approx 1.16\times 10^{-13} \left(\frac{k}{k_\lambda}\right)^{0.2}.
\ee
Since data will only constrain $P_{B_0}$ directly, we take this
equation to be exact for our numerical results so they may easily be
related to different power spectra for $P_{B_0}$ (which may or may not
come from the assumed power law spectrum of $B_a$ fluctuations). The power
spectrum $P_{B_0}$ scales as $B_\lambda^4$ as do the CMB power
spectra, so large changes in overall amplitude can be obtained from
relatively small changes in the primordial field: the value of
$B_\lambda$ has to be quite finely tuned to give a CMB signature that
is neither totally dominant nor totally negligible.

Numerical results from vector modes with $B_\lambda = 3\times
10^{-9}\Gauss$ are shown in
Fig.~\ref{vecCls}, in
comparison with the spectra expected from primordial
curvature perturbations and possible primordial gravitational waves. For this $B_\lambda$ the effect on the temperature power
spectrum is negligible; only if $B_\lambda \agt 8\times 10^{-9}\Gauss$
could there be a significant contribution to the power at $l\agt
2000$, perhaps contributing some of the power observed on these scales~\cite{Readhead:2004gy}.

The contributions to the most easily distinguished B-modes
are shown in Fig.~\ref{BCls}, including the tensor contribution. It is
clear that the compensated tensor mode has a negligible observational
signature and can be neglected. The exact amplitude of the large scale
tensor signal from gravitational
waves sourced before neutrino decoupling is uncertain because we do not know
the time (or mechanism) of field generation, nor have we modelled neutrino decoupling
in detail.

Our vector mode results are in broad agreement with the semi-analytical results
of~\cite{Subramanian:2003sh}. The main qualitative difference is that
our $TE$ cross-correlation changes sign in the damping region. The quantitative
results differ somewhat across the spectrum due to our more detailed
full analysis of the damping, recombination, inclusion of neutrinos
and modelling of reionization (we have also used a slightly different
primordial power spectrum). The results in Ref.~\cite{Mack:2001gc} for the large scale vector and
tensor spectra are too large by a factor\footnote{An inconsistency between their definition Eq. 2.17
and the equation for the $C_l$, Eq. 5.1} of
$(2\pi)^3$ (giving constraints on $B_\lambda$ too small by a factor of
$(2\pi)^{3/4} \sim 4$). There was another normalization error 
in~\cite{Subramanian:2002nh} but corrected in~\cite{Subramanian:2003sh}.
Refs.~\cite{Mack:2001gc,Durrer:1999bk} provide
analytical solutions valid for tensor modes that are super-horizon at
decoupling, which give $C_l$ spectra qualitatively valid for $\ell \alt 100$. However this
approximation was also used for $l < 500$ in Ref.~\cite{Mack:2001gc}, and so their result is
qualitatively incorrect at $l \agt 100$. Their
tensor polarization results also suffer qualitative problems because
the peak in the visibility was neglected. All previous analyses of sourced
tensor modes have neglected neutrino compensation, giving results somewhat
larger (but the result is still, in any case, somewhat uncertain).

\section{Conclusions}

Nearly scale invariant primordial magnetic fields can give a
potentially observable CMB signal if they happen to be $\agt
10^{-10}\Gauss$.  The observational B-mode
signature comes from tensor modes, giving a non-Gaussian spectrum otherwise
essentially identical to that expected from primordial gravitational
waves, and vector modes giving power on small scales. 

Any possible future detection of primordial gravitational waves from
large scale B-modes should be carefully distinguished from that
produced by magnetic fields. Any primordial signal is
expected to be Gaussian, so Gaussianity tests can be used to
distinguish them (methods for robustly isolating the B-mode component on sections of the
sky are given in Ref.~\cite{Lewis:2003an}, and can be used to
construct a set of cut-sky modes that should be Gaussian if they are
due to inflation).  Low frequency observations may also be able to
detect Faraday rotation~\cite{Kosowsky:1996yc}, which would
be a clear signal of magnetic fields.
Small scale B-mode observations from magnetic fields will need to be
carefully distinguished from the weak lensing signal. Regular
primordial vector modes, though theoretically unmotivated,  can also in principle give a significant
small and large scale B-mode signal~\cite{Lewis:2004kg}. They may be
distinguished by their sharper fall in power on very small scales due
to the lack of sources. Topological defects~\cite{Seljak:1997ii} can be identified by the lack of non-Gaussian
tensor mode power
on large scales. Note that throughout we have been assuming idealized
observations; in practice foregrounds and systematics may well pose very
serious problems (see e.g. Ref.~\cite{Tucci:2003wd}).

Our analysis is significantly more detailed
than previous work, in that we have numerically solved the full set of
linearized equations. There is a qualitatively important
mechanism of a neutrino anisotropic stress compensation on
super-horizon scales that was previously neglected. Computing the full transfer functions is rather
straightforward, and we encourage future workers in this area to at least
compare their semi-analytic results with the numerical answer to
ensure that important physical effects have not be accidentally
overlooked. Our modified version of CAMB\footnote{\url{http://camb.info}}~\cite{Lewis:1999bs} for efficiently computing vector mode power
spectra is publicly available, and may also be useful for
computing anisotropies from other sources, for example topological
defects or second order effects.

\section*{Acknowledgments}
I thank Marco Peloso, Anthony Challinor, Arthur Kosowsky, Tinatin
Kahniashvili and Christos Tsagas for useful communication.

\appendix

\newpage
\begin{widetext}
\section{Multipole equations, harmonic expansion and $C_l$}

In this appendix we review in a streamlined fashion the multipole
equations, solutions, and equations for the $C_l$ needed for numerical
calculation~\cite{Challinor:1998xk,Challinor:1999xz,Challinor:2000as}.
The definitions used here are precisely those used in the CAMB~\cite{Lewis:1999bs} numerical implementation.
Equivalent results using the total angular momentum method are
given in Ref.~\cite{Hu:1997hp}.

The photon multipole evolution is governed by the geodesic equation
and Thomson scattering, giving~\cite{Challinor:2000as}

\begin{multline}
\dot{I}_{A_l} + \frac{4}{3}\Theta I_{A_l} + D^bI_{b A_l} -
\frac{l}{2l+1} D_{\la a} I_{A_{l-1}\ra} + \frac{4}{3} I
A_{a_1} \delta_{l1} - \frac{8}{15} I \sigma_{a_1
  a_2}\delta_{l2}\\
= -n_e\sigma_T\left( I_{A_l} - I\delta_{l0} - \frac{4}{3} I
  v_{a_1}\delta_{l1} -\frac{2}{15}\zeta_{a_1 a_2} \delta_{l2}\right)
\label{IAmult}
\end{multline}
where $I_{A_l}$ is taken to be zero for $l<0$ and 
\be
\zeta_{ab} \equiv \frac{3}{4} I_{ab} + \frac{9}{2} \calE_{ab}
\ee
is a source from the anisotropic stress and E-polarization. The 
equation for the density perturbation $D_a I$ is obtained by taking the spatial
derivative of the above equation for $l=0$.
The corresponding evolution equations for the polarization multipole
tensors are~\cite{Challinor:2000as}
\ba
\dot{\calE}_{A_l} + \frac{4}{3}\Theta \calE_{A_l} +
\frac{(l+3)(l-1)}{(l+1)^2} \uD^b\calE_{b A_l} - 
\frac{l}{2l+1} D_{\la
a_l} \calE_{A_{l-1}\ra} - \frac{2}{l+1} \curl \calB_{A_l} &=& -n_e
\sigma_T(\calE_{A_l} - \frac{2}{15} \zeta_{a_1 a_2} \delta_{l2}) \nonumber\\
\dot{\calB}_{A_l} + \frac{4}{3}\Theta \calB_{A_l} +
\frac{(l+3)(l-1)}{(l+1)^2} \uD^b\calB_{b A_l} - 
\frac{l}{2l+1} D_{\la
a_l} \calB_{A_{l-1}\ra} + \frac{2}{l+1} \curl \calE_{A_l} &=& 0.
\label{EBAmult}
\ea
\end{widetext}

For numerical solution we expand the covariant equations into
scalar, vector and tensor harmonics. The resulting equations for the
modes at each wavenumber can be studied easily and also integrated numerically.

\subsection{Scalar, vector, tensor decomposition}

It is useful to do a decomposition into $m$-type tensors, scalar
($m=0$), vector ($m=1$) and 2-tensor ($m=2$) modes. They describe
respectively density perturbations, vorticity and gravitational waves.
In general
a rank$-\ell$ PSTF tensor $X_{A_l}$ can be written as a sum of
$m-$type tensors
\ba
X_{A_l} = \sum_{m=0}^l X^{(m)}_{A_l}.
\ea
Each $X^{(m)}_{A_l}$ can be written in terms of $l-m$ derivatives of a
transverse tensor
\be
X^{(m)}_{A_l} = D_{\la A_{l-m}} \Sigma_{A_m \ra}
\ee
where $\uD_{A_l} \equiv \uD_{ a_1}  \uD_{a_2} \dots \uD_{a_l}$
and $\Sigma_{A_m}$ is first order, PSTF and transverse $\uD^{a_m}
\Sigma^m_{A_{m-1}a_m} = 0$.  The `scalar' component is $X^{(0)}$, the `vector' component
is $X^{(1)}_a$, etc. Since General Relativity gives no sources for $X_{A_m}$ with
$m>2$ usually only scalars, vectors and (2-)tensors are considered. At
linear order they evolve independently.

\subsection{Harmonic expansion}

For numerical work we perform a harmonic expansion in terms of
zero order eigenfunctions of the Laplacian $Q^m_{A_m}$, 
\be
\uD^2 Q^m_{A_m} = \frac{k^2}{S^2} Q^m_{A_m},
\ee
where $Q^m_{A_m}$ is transverse on all its indices, $\uD^{a_m}Q^m_{A_{m-1}a_m} = 0$.
So a scalar would be expanded in terms of $Q^0$, vectors in terms of
$Q^1_a$, etc. We usually suppress the labelling of the different
harmonics with the same eigenvalue, but when a function depends only
on the eigenvalue we write the argument explicitly, e.g. $f(k)$. 

For $m>0$ there are eigenfunctions with positive and negative parity, which we
can write explicitly as $Q_{A_m}^{m\pm}$ when required. Since
\be
\uD^2 (\curl Q_{A_m}) = \curl (\uD^2 Q_{A_m}) = \frac{k^2}{S^2} \curl Q_{A_m}
\ee
they are related by the curl operation. Using the result
\be
\curl\curl Q^m_{A_m} =
\frac{k^2}{S^2}\nonflat{\left[1+(m+1)\frac{K}{k^2}\right]} Q^m_{A_m}
\ee
we can choose to normalize the $\pm$ harmonics the same way so that
\be
\curl Q^{m\pm}_{A_m} = \frac{k}{S}\nonflat{\sqrt{1+(m+1)\frac{K}{k^2}}} Q^{m\mp}_{A_m}.
\ee
A rank-$\ell$ PSTF tensor of either parity may
be constructed from $Q^{m\pm}_{A_m}$ as
\ba
Q^m_{A_l} \equiv \left(\frac{S}{k}\right)^{l-m} \uD_{\la A_{l-m}} Q^m_{A_m \ra}
\ea
and an $X^{(m)}_{A_l}$ component of $X_{A_l}$ may be expanded in terms of
these tensors. They satisfy
\ba
\uD^2 Q^m_{A_l} &=& \frac{k^2}{S^2}\nonflat{\left( 1 -
    [l(l+1)-m(m+1)]\frac{K}{k^2}\right)} Q^m_{A_l}\nonumber\\
\uD^{a_l} Q^m_{A_{l-1}a_l} &=&
\nonflat{\kf_l^m} 
\frac{k}{S} \frac{(l^2-m^2)}{l(2l-1)} Q^m_{A_{l-1}}\nonumber
\\
\curl Q_{A_l}^{m \pm}&=& \nonflat{\sqrt{\kf_0^m}} 
\frac{m}{l} \frac{k}{S}  Q^{m \mp}_{A_l} 
\label{Qident}
\ea
where
\nonflat{ $\kf_l^m \equiv 1 - \left\{l^2 - (m+1)\right\}K/k^2$ and}
$l\ge m$. 

Dimensionless harmonic coefficients are defined by
\begin{gather}
\begin{aligned}
\sigma_{ab}^{(m)} &= \sum_{k,\pm} \frac{k}{S} \sigma^{(m)\pm} Q^{m\pm}_{ab}
&
H_{ab}^{(m)} &= \sum_{k,\pm} \frac{k^2}{S^2} H^{(m)\pm}  Q^{m\pm}_{ab}
\nonumber\\
q_a^{(m)} &= \sum_{k,\pm} q^{(m)\pm}  Q_a^{m\pm}
&
E_{ab}^{(m)} & = \sum_{k,\pm} \frac{k^2}{S^2} E^{(m)\pm} Q_{ab}^{m\pm}
\nonumber\\
\pi_{ab}^{(m)} &= \sum_{k,\pm} \Pi^{(m)\pm} Q_{ab}^{m\pm}
&
I_{A_l}^{(m)}& =\rho_\gamma \sum_{k,\pm} I_l^{(m)\pm} Q^{m\pm}_{A_l}\nonumber\\
\Omega_a &= \sum_{k,\pm} \frac{k}{S} \Omega^{\pm} Q^{1\pm}_a
&
 A_a^{(m)} &= \sum_{k,\pm} \frac{k}{S} A^{(m)\pm} Q^{m\pm}_a \nonumber
\end{aligned}
\\(D_a X)^{(m)}  =  \sum_{k,\pm} \frac{k}{S} (\delta X)^{(m)}{}^\pm Q^{m\pm}_a
\label{Qexpand}
\end{gather}

where the $k$ dependence of the harmonic coefficients is
suppressed. We also often suppress $m$ and $\pm$ indices for clarity.
The other multipoles are expanded in analogy with $I_{A_l}$.
The heat flux for each fluid component is given by 
$q_i=(\rho_i+p_i) v_i$, where
$v_i$ is the velocity, and the total heat flux is given by
$q=\sum_i q_i$.  We write the baryon velocity simply as $v$, and
define a constant $B_0^{(m)} \equiv \Pi^{(m)}_B/\rho_\gamma$ to
quantify the magnetic field anisotropic stress source. In the frame in
which $\Omega_a = 0$ gradients are purely scalar $\bar{(\delta X)}^{(1)} = 0$.

\subsection{Harmonic multipole equations}

Expanded into harmonics, the photon multipole equations~\eqref{IAmult} become
\begin{multline}
I_l' + \frac{k}{2l+1}\left[ 
\nonflat{\kf_{l+1}^m} 
\frac{(l+1)^2-m^2}{l+1} I_{l+1} - l
  I_{l-1}\right] =\\
-S n_e\sigma_T\left( I_l -\delta_{l0}I_0 - \frac{4}{3}\delta_{l1}v -
\frac{2}{15}\zeta \delta_{l2}\right) \\+ \frac{8}{15} k \sigma \delta_{l2} -
4h'\delta_{l0} - \frac{4}{3}k A \delta_{l1} 
\label{Imult}
\end{multline}
where $l\ge m$, $I_0=\delta\rho_\gamma/\rho_\gamma$, $I_l=0$ for
  $l<m$, and $m$ superscripts are implicit. The scalar source is $h'=(\delta S/S)'$.
The equation for the neutrino multipoles (after neutrino decoupling) is the same but without the
  Thomson scattering terms (for massive neutrinos see Ref.~\cite{Lewis:2002nc}).
The polarization multipole equations~\eqref{EBAmult} become
\begin{widetext}
\begin{align}
\calE^{m\pm}_l{}' + 
k\left[\nonflat{\kf_{l+1}^m} 
\frac{(l+3)(l-1)}{(l+1)^3} \frac{(l+1)^2-m^2}{(2l+1)}
\calE_{l+1}^{m\pm}
 - \frac{l}{2l+1}  \calE_{l-1}^{m\pm} - \frac{2m}{l(l+1)}\nonflat{\sqrt{\kf_0^m}}  \calB^{m\mp}_l\right] &= -Sn_e\sigma_T(\calE_l^{m\pm} - \frac{2}{15} \zeta^{m\pm}\delta_{l2})\nonumber\\
\calB^{m\pm}_l{}' + k\left[ 
\nonflat{\kf_{l+1}^m} 
\frac{(l+3)(l-1)}{(l+1)^3} \frac{(l+1)^2-m^2}{(2l+1)}
\calB_{l+1}^{m\pm}
 - \frac{l}{2l+1}  \calB_{l-1}^{m\pm} + \frac{2m}{l(l+1)}
 \nonflat{\sqrt{\kf_0^m}} \calE^{m\mp}_l\right] &= 0.
\label{EBmult}
\end{align}

\subsection{Integral solutions}

Solutions to the Boltzmann hierarchies can be found in terms of line of
sight integrals. 
\nonflat{For simplicity we only give results for a flat universe here.}
The $I_l^m$ hierarchy has homogeneous solutions
(i.e. solutions to Eq.~\ref{Imult} with RHS set to zero) given by derivatives of
\ba
\Psi_l^m(k\eta) \equiv \frac{l!}{(l-m)!} \frac{j_l(k\eta)}{(k\eta)^m}
\ea
where $j_l(x)$ is a spherical Bessel function.
These can be used to construct Green's function solutions to the full equations.
For the polarization the result is less obvious, 
 though solutions can
easily be verified once found.
For vector modes ($m=1$) the solutions are~\cite{Lewis:2004kg}
\ba
I_l(\eta_0) &=& 4\int^{\eta_0} \!\!\!d\eta e^{-\tau}\biggl[ Sn_e \sigma_T \bv
\Psil^1(\chi)
 + \left(k \bsigma + Sn_e\sigma_T \frac{\zeta}{4} \right)
\frac{d\Psil^1(\chi)}{d\chi}\biggr]\label{vecintsol} \\
E_l^\pm(\eta_0) &=&
 \frac{l(l-1)}{l+1} \int^{\eta_0} d\eta S n_e \sigma_T
e^{-\tau}\left[ \frac{1}{\chi}\frac{d j_l(\chi)}{d\chi} + \frac{j_l(\chi)}{\chi^2}\right] \zeta^\pm
\\
B_l^\pm(\eta_0) &=& -\frac{l(l-1)}{l+1} \int^{\eta_0} d\eta S n_e \sigma_T
e^{-\tau} \frac{j_l(\chi)}{\chi} \zeta^\mp
\ea
where $\chi \equiv k(\eta_0-\eta)$. For tensors ($m=2$) the solutions
are~\cite{Challinor:2000as}
\ba
I_l(\eta_0) &=& 4\int^{\eta_0} \!\!\!d\eta e^{-\tau} \left[ k\sigma +
  S n_e \sigma_T \frac{\zeta}{4} \right] \Psil^2(\chi) 
\\
E_l^\pm(\eta_0) &=&
 \frac{l(l-1)}{(l+1)(l+2)}\int^{\eta_0} \!\!\!d\eta Sn_e \sigma_T e^{-\tau}\left[ \frac{d^2 j_l(\chi)}{d\chi^2}  +
   \frac{4}{\chi} \frac{d j_l(\chi)}{d\chi} - \left(1 -
     \frac{2}{\chi^2}\right) j_l(\chi) \right]\zeta^\pm \\
B_l^\pm(\eta_0) &=&
 -2\frac{l(l-1)}{(l+1)(l+2)}\int^{\eta_0} \!\!\!d\eta Sn_e \sigma_T e^{-\tau}\left[ \frac{d j_l(\chi)}{d\chi} +
   \frac{2}{\chi} j_l(\chi)\right]\zeta^\mp.
\ea
Here $\tau$ is the optical depth from $\eta$ to $\eta_0$, $\tau' = -S
n_e \sigma_T$.
\subsection{Power spectra}
Using the harmonic expansion of $I_{A_l}$ in Eq.~\ref{C_l} the
contribution to the $C_l$ from
type-$m$ tensors becomes
\be
C^{TT(m)}_l = 
\frac{\pi }{4}
\frac{(2l)!}{(-2)^l(l!)^2} 
\sum_{k,k',\pm}  \la I^\pm_{l,k} I^\pm_{l,k'} \ra Q_{A_l k}^\pm Q^{A_l\pm}_{k'}.
\ee
The multipoles $I_l$ can be related to some primordial variable  $X_{A_m} = \sum_k \left( X^+ Q^{m+}_{A_m} +  X^- Q^{m-}_{A_m}\right)$ 
via a transfer function $T^X_l$ defined by $I_l = T_l^X X$.
Statistical isotropy and orthogonality of the harmonics implies that
\be
 \la X^\pm_k X^\pm_{k'} \ra = f_X(k) \delta_{kk'}
\ee
where $\sum_k \delta_{kk'} Y_k  = Y_{k'}$ and $f_X(k)$ is some
function of the eigenvalue $k$. The normalization of the $Q_{A_l}^m$
is given by 
\be
 \int dV Q_{A_l}^m Q^m{}^{A_l}
=  \int dV
\left(\frac{-S}{k}\right)^{l-m}  Q_{A_m}^m D^{A_{l-m}}
 Q^m_{A_l}=
\nonflat{\kfprod_l^m} \frac{(-2)^{l-m} (l+m)!(l-m)!}{(2l)!}  N
\ee
where we have integrated by parts repeatedly, then 
repeatedly applied the identity for
the divergence~\eqref{Qident}. 
The normalization is $N\equiv \int dV Q^{A_m} Q_{A_m}$\nonflat{and $\kfprod^l_m \equiv \prod_{n=m+1}^l \kf^m_n$}.
By statistical isotropy $C_l = (1/V) \int dV C_l$
and hence
\be
C^{TT(m)}_l = 
\frac{\pi }{4}
\frac{(l+m)!(l-m)!}{(-2)^m(l!)^2}
 \sum_{k,\pm} \frac{N}{V}\nonflat{\kfprod_l^m} |T_l^X(k)|^2 f_X(k)
\ee

We choose to define a power spectrum $P_X(k)$ so that
the real space isotropic variance is given by 
\be
\la |X_{A_m} X^{A_m}| \ra = \sum_{k,\pm} \frac{|N|}{V} f_X(k)  \equiv \int d \ln k \,\,  P_X(k)
\label{Pdef}
\ee
so the CMB power spectrum becomes
\be
C^{TT(m)}_l = 
\frac{\pi }{4}
\frac{(l+m)!(l-m)!}{2^m (l!)^2} \int d\ln k\,\, P_X(k) \nonflat{\kfprod_l^m}|T_l^X(k)|^2.
\ee
\nonflat{For a non-flat universe $\int d\ln k$ is replaced by some
  some other measure which should be specified when defining the power
  spectrum. In a closed universe the integral becomes a 
  sum over the discrete modes.}
Note that we have not had to choose a specific representation of
$Q_{A_m}$ or $\sum_k$.

The polarization $C_l$ are obtained similarly~\cite{Challinor:2000as}
and in general we have
\ba
C^{JK(m)}_l &=& 
\frac{\pi}{4}\left[\frac{(l+1)(l+2)}{l(l-1)}\right]^{p/2} \frac{(2l)!}{(-2)^l
  (l!)^2} \frac{ \la J_{A_l} K^{A_l} \ra}{\rho_\gamma^2} \\
&=& 
\frac{\pi}{4} \left[\frac{(l+1)(l+2)}{l(l-1)}\right]^{p/2}
\frac{(l+m)!(l-m)!}{2^m(l!)^2} \int d\ln k \,\, P_X(k)\,
\nonflat{\kfprod_l^m} J^X_l K^X_l
\ea
\end{widetext}
where $JK$ is $TT$ ($p=0$), $EE$ or $BB$ ($p=2$) or $TE$ ($p=1$). Our
conventions for the polarization are consistent with
CMBFAST~\cite{Seljak:1997gy} and
CAMB~\cite{Lewis:1999bs}. We have assumed a parity symmetric ensemble,
so $C_l^{TB}=C_l^{EB}=0$.

For tensors we use $H_T$ where $h_{ij} = \sum_{k,\pm} 2H_T
Q^2_{ij}$ and $h_{ij}$ is the transverse traceless part of the metric
tensor. This introduces an additional factor of $1/4$ into the
result for the $C_l$ in terms of $P_h$ and $T^{H_T}_l$.

The numerical factors in the hierarchy and $C_l$ equations depend on
the choice of normalization for the $\ell$ and $k$ expansions. Neither
$e_{\la A_l \ra}$ nor $Q_{A_l}$ are normalized, so there are
compensating numerical factors in the expression for the $C_l$. If
desired one can do normalized expansions, corresponding to an
$\ell$-\nonflat{ and $k$-}dependent
re-scaling of the $I_l$ and other harmonic coefficients, giving
expressions in more manifest agreement with Ref.~\cite{Hu:1997hp}.


\end{document}